\documentclass[12pt]{article}
\usepackage{amsmath,amssymb}
\usepackage{graphicx,color}
\numberwithin{equation}{section}
\usepackage{cite}
\usepackage{bm}
\usepackage{dcolumn}
\newcommand{\be}{\begin{equation}}
\newcommand{\bea}{\begin{eqnarray}}
\newcommand{\eea}{\end{eqnarray}}
\newcommand{\ba}{\begin{array}}
\newcommand{\ea}{\end{array}}
\newcommand{\ee}{\end{equation}}

\expandafter\ifx\csname mathbbm\endcsname\relax

\else

\fi \textheight 22cm \textwidth 15cm \topmargin 1mm
\begin{document}

\begin{titlepage}
\hfill
\vbox{
    \halign{#\hfil         \cr
           IPM/P-2008/029 \cr
                      } 
      }  
\vspace*{20mm}
\begin{center}
{\Large {\bf Central Charge for 2D Gravity on $AdS_2$ and $AdS_2/CFT_1$ Correspondence}\\
}

\vspace*{15mm}
\vspace*{1mm}
{Mohsen Alishahiha$^a$ and  Farhad Ardalan$^{a,b}$}

 \vspace*{1cm}

{\it ${}^a$ School of physics, Institute for Research in Fundamental Sciences (IPM)\\
P.O. Box 19395-5531, Tehran, Iran \\ }

\vspace*{.4cm}

{\it ${}^b$ Department of Physics, Sharif University of Technology \\
P.O. Box 11365-9161, Tehran, Iran}

\vspace*{2cm}
\end{center}

\begin{abstract}

We study 2D Maxwell-dilaton gravity on $AdS_2$. We distinguish two distinctive cases 
depending on whether the $AdS_2$ solution can be lifted to an $AdS_3$ geometry.
In both cases, in order to get a consistent boundary condition we need to work with 
a twisted energy momentum tensor which has non-zero central charge. With this 
central charge and the explicit form of the twisted Virasoro generators we compute the 
entropy of the system using the Cardy formula. 
The entropy is found to be the same as that obtained from gravity calculations for a
specific value of the level of the $U(1)$ current.
The agreement is an indication of $AdS_2/CFT_1$
correspondence.
\end{abstract}
\end{titlepage}

\section{Introduction}

In 3D gravity the group of diffeomorphisms which preserves the condition that the metric be 
 asymptotically $AdS_3$ is two copies of the Virasoro algebra with the central charge\cite{BH}
\be
c=\frac{3l_3}{2G_3},
\label{c}
\ee
where $l_3$ is the $AdS_3$ radius and $G_3$ is the three dimensional Newton's constant.

This fact has been used to compute the entropy of three dimensional black holes. It has been 
shown \cite{Strominger:1997eq} that one may use the Cardy formula with 
the above central charge
and count the boundary degrees of freedom which agrees with the entropy of the black hole in the bulk.
Therefore the symmetry is enough to find the entropy without knowing about the details 
of the dynamics.

Of course we now understand the reason behind this precise agreement and it is due to 
the  AdS/CFT correspondence \cite{Maldacena:1997re}. In three dimensions the lesson we have
learned from the AdS/CFT correspondence is that 
``any consistent quantum gravity on an asymptotically $AdS_3$ spacetime is a 2D CFT living 
on the boundary of the $AdS_3$''. 

Although AdS$_{d+1}$/CFT$_{d}$ have been understood for $d\geq 2$ mainly due to explicit examples, 
little has been known for the case of $d=1$ (see however \cite{{Strominger:1998yg},{Cadoni:1999ja},{NavarroSalas:1999up},{Cadoni:2000ah},{Cadoni:2000gm},
{Sen:2008yk}}). 
The lack of our knowledge
of the holographic dual of $AdS_2$ is due to the special features of $AdS_2$ spacetime. First
of all it has two boundaries. 
Secondly, so far we do not have a concrete 
example of an AdS$_2$/CFT$_1$ in the context of string theory where we could identify both
sides of the duality. 

On the other hand the quantum gravity on $AdS_2$ geometry is important on its own right. Indeed the $AdS_2$ geometry 
is the factor which appears in the near horizon geometry of the extremal black holes in any dimension. Therefore
understanding gravity on $AdS_2$ might ultimately help us understand the origin of the black hole
entropy in other dimensions.

To explore the AdS$_2$/CFT$_1$ correspondence one may utilize the experience of the $AdS_3$ case;
namely, one could try to understand the asymptotic symmetry of $AdS_2$. In fact this has been done
in several papers including \cite{{Strominger:1998yg},{Cadoni:1999ja},
{NavarroSalas:1999up},{Cadoni:2000ah}}. In particular it has been shown \cite{Strominger:1998yg}
that the asymptotic symmetry group of an asymptotically $AdS_2$ geometry is one copy of the Virasoro 
algebra. Recently
it has also been shown \cite{HS} that exactly the same argument as that for $AdS_3$ \cite{BH}
can be made for quantum gravity with a $U(1)$ gauge field on $AdS_2$ leading to the following 
central charge\footnote{In order to compare the result with that in \cite{HS} one needs to use a unit in which
$G=\frac{1}{4}$.}
\be
c=12kG^2Q^2l^4,
\label{cc}
\ee
where $l$ is the radius of $AdS_2$, Q is the electric charge, $k$ is the level of the 
current which generates the $U(1)$ and $G$ is the two dimensional Newton's constant.

The aim of this article is to further study 2D Maxwell-dilaton gravity on $AdS_2$ background.
In particular we would like to study the entropy of the corresponding solution and to see
to what extent the information of the theory is encoded in a CFT whose central charge is given by
(\ref{cc}). Our strategy is the 2D analog of \cite{Strominger:1997eq}. Namely we will
reproduce the black hole entropy obtained from gravity by making use of the Cardy formula with 
the central charge given by the asymptotic conformal diffeomorphism of the 2D theory.
As observed in \cite{HS} in order to have a consistent boundary condition 
for gravity on $AdS_2$ coupled to gauge field, the 
energy momentum tensor has to be twisted. 
Having a non-zero central charge plus the fact that we can consistently reproduce the entropy from this
central charge provides an indication of the
$AdS_2/CFT_1$ correspondence showing that the corresponding holographic dual would be 
a chiral half of a 2D CFT (see also \cite{{HS},{Astorino:2002bj}}).

In the course of studying 2D Maxwell-dilaton gravity on $AdS_2$ we will encounter two distinctive cases 
depending on whether the corresponding $AdS_2$ solution 
can be lifted up into $AdS_3$ solution. These two models are given by
the actions
\bea\label{actions}
S_1&=&\frac{1}{8G}\int d^2x\sqrt{-g} \left(e^{\phi}(R+\frac{8}{l^2})-\frac{l^2}{4}F^2\right),\cr &&\\
S_2&=&\frac{1}{8G}\int d^2x\sqrt{-g}\; e^{\phi}\left(R+\frac{2}{l^2}-\frac{l^2}{4}e^{2\phi}F^2\right).
\nonumber
\eea
Both actions admit an $AdS_2$ vacuum solution. We note, however, that although in the second case 
the $AdS_2$ solution can be lifted up to three dimensional $AdS_3$ solution, in the first one it cannot.
Our main conclusion is that for both cases
we can reproduce the black hole entropy using the central charge of the twisted energy 
momentum tensor.
Moreover we can show that in the first case approaching the horizon a CFT emerges which
has the same central charge as that obtained by using the asymptotic symmetry. Therefore
in this case there may be a correspondence between the two CFT's;
one at infinity and the other at the horizon. 

The paper is organized as follows. In section two we will study 2D Maxwell-dilaton 
gravity on $AdS_2$ based on the first action in (\ref{actions}). We will show that 
the central charge of the twisted energy momentum can be used to compute the entropy
using the Cardy formula. The entropy is the same as the black hole entropy obtained from
the gravity side for a specific value of $k$. Therefore the consistency of the
results leads us to fix the 
level of $U(1)$ current too. We will also study the model in terms of the near horizon modes
where we show that at near horizon we get a CFT whose central charge is the same 
at that obtained from asymptotic symmetry. In section three we will do similar computations 
for the model based on the second action in
(\ref{actions}). We will see that using the asymptotic symmetry one can read off the central 
charge of the twisted energy momentum tensor which can then be used to reproduce the black hole entropy 
correctly. The last section is devoted to the conclusions and
discussions.

\section{Type I 2D gravity}

\subsection{Central charge and entropy}

In this section we study the 2D Maxwell-dilaton gravity based on the action  
\be
S_1=\frac{1}{8G}\int d^2x\sqrt{-g} \left(e^{\phi}(R+\frac{8}{l^2})-\frac{l^2}{4}F^2\right).
\ee
This model has recently been considered in \cite{HS} where the central charge of the CFT
associated with the asymptotic symmetry of its $AdS_2$ solution was calculated. 
The authors of \cite{HS} 
noticed that the potential of the $U(1)$
gauge field is singular at the boundary and therefore the boundary condition is not respected 
under a conformal diffeomorphism. The resolution was to accompany the conformal transformation with  
a special $U(1)$ gauge transformation, generated by a 
current $j_{\pm}$, leading to the twisted energy momentum tensor\cite{HS}
\be
\tilde{T}_{\pm\pm}=T_{\pm\pm}\pm {GQl^2}\partial_\pm j_\pm,
\label{ee}
\ee
where T is the energy momentum tensor corresponding to the original conformal transformation with $c=0$. 

Since the $AdS_2$ solution carries an entropy, it is natural to pose the question of whether
this entropy can be reproduced using  the Cardy formula with central charge given by
(\ref{cc}). In other words we should be able to compute the {\it statistical} entropy  
by evaluating the eigenvalue of $\tilde{L}_0$ coming from energy momentum (\ref{ee}) 
and then use the Cardy formula with the central charge (\ref{cc}). The resultant 
entropy should be compared with
the entropy of the $AdS_2$ solution.

Therefore it is important to first compute the entropy of the model using the gravity solution
which we do utilizing the entropy function formalism \cite{SEN}, which only needs
the information of the near horizon geometry. Since the geometry is $AdS_2$, 
we start from an ansatz respecting the $SO(2,1)$ isometry
of the $AdS_2$ solution,
\be
ds^2=v\left(-r^2 dt^2+\frac{dr^2}{r^2}\right),\;\;\;\;\;e^{\phi}=\eta,\;\;\;\;\;
F_{rt}=\frac{e}{l^2},
\ee
where $v,\eta$ and $e$ are constant to be determined by the equations of motion. 

To proceed we need to evaluate the entropy function,  
\be
{\cal E}=2\pi (Qe-f(e,v,\eta)),
\ee
where 
\be 
f=\frac{v}{8G} \left[\eta\left(-\frac{2}{v}+\frac{8}{l^2}\right)+\frac{e^2}{2v^2l^2}\right]
\ee
is the Lagrangian density evaluated for the ansatz. Extremizing the
entropy function with respect to parameters $v,\eta$ and $e$ we get
\be
ds^2=\frac{l^2}{4}\left(-r^2 dt^2+\frac{dr^2}{r^2}\right),\;\;\;\;\;e^{\phi}=4G^2Q^2l^4,\;\;\;\;\;
F_{rt}=2GQl^2.
\ee
The entropy is also given by 
\be
S_{BH}=2\pi GQ^2l^4
\label{BHE}
\ee 
which can be recast in the suggestive form
\be
S_{BH}=2\pi \sqrt{\frac{1}{6}(12GQ^2l^4)(\frac{12GQ^2l^4}{24})},
\ee 
reminiscent  of the Cardy formula 
\be
S=2\pi \sqrt{(\frac{c}{6}-4{\Delta}_0)({\Delta}-\frac{c}{24})},
\label{cardy}
\ee
where $\Delta$ is the eigenvalue of $L_0$ and $\Delta_0$ its the lowest eigenvalue. 
With this observation, the task is to see
whether the CFT defined by the twisted energy momentum tensor (\ref{ee}) and central charge 
(\ref{cc}) can in fact reproduce the above entropy.

To proceed we note that the twisted energy momentum tensor (\ref{ee}) satisfies the following 
commutator relation \cite{HS}
\bea
[\tilde{T}_{--}(t^-),\tilde{T}_{--}(s^-)]&=&-4\pi\partial_-\delta(t^--s^-)\tilde{T}_{--}(s^-)
+2\pi\delta(t^--s^-)\partial_-\tilde{T}_{--}(s^-)\cr &+&2\pi k G^2Q^2l^4\partial_-^3\delta(t^--s^-),
\eea
leading to the central charge of $c=12kG^2Q^2l^4$. 
$k$ is the coefficient of the Schwinger term in the $U(1)$ current algebra
\be
[j_-(t^-),j_-(s^-)]=-2\pi k\partial_-\delta(t^--s^-).
\label{jj}
\ee
These expressions together with the definition of the twisted energy momentum tensor show that
upon mode expanding the energy momentum tensor, the Virasoro generators become  
\be
\tilde{L}_n=L_n-GQl^2U_n,
\ee
which satisfy a Virasoro algebra as follows 
\be
[\tilde{L}_n,\tilde{L}_m]=(n-m)\tilde{L}_{n+m}+\frac{12kG^2Q^2l^4}{12}(n^3-n)\delta_{n+m,0}.
\ee
Here $U_n$ are the coefficients of  the mode expansion of $\partial_-j_-$. 

We now want to evaluate the eigenvalues of $L_0$ and $U_0$. This may be done by using the 
semiclassical method in which $\Delta$ is obtained from the integral of the energy momentum tensor
$T$ evaluated at the $AdS$ solution which turns out to be zero. On the other hand to find 
$U_0$  note that the non-zero value of the divergence of the $U(1)$ current, $\partial_-j_-$,
which gives an important contribution to the energy momentum tensor, is 
the anomaly related to the Schwinger term in (\ref{jj}). 
As it has already been pointed out in \cite{HS} the situation is very similar
to the Schwinger model \cite{Schwinger:1962tp} where, using the fermionic formulation,
one gets the central term in the commutator, and the anomaly in the $U(1)$ current, due to the one
loop digram contributions.  
This observation can be used to evaluate the divergence of the current.
Indeed using the standard anomaly  and the Schwinger term calculations one finds 
(see for example \cite{{Manton:1985jm},{Heinzl:1991vd}})
\be
\partial_-j_-=\frac{k}{4}\;\epsilon^{\mu\nu}F_{\mu\nu}=-kGQl^2.
\ee
So that $\tilde{\Delta}=kG^2Q^2l^4$.
Therefore with the assumption $\tilde{\Delta}_0=0$ and using the central charge  
$c=12kG^2Q^2l^4$ we arrive at
\be
S=2\pi G Q^2l^4 \;(kG).
\ee
Comparing the Cardy entropy with the black hole entropy (\ref{BHE}) one has to set $k=\frac{1}{G}$. 
Thus to get a consistent result the level of $U(1)$ is not 
a free parameter. Of course we are familiar with such a phenomena; namely requiring to 
get correct black hole entropy may put a constraint on the level of affine algebra in the
dual CFT (for example see \cite{{Giveon:2006pr},{Dabholkar:2007gp},
{Johnson:2007du},{Lapan:2007jx},{Kraus:2007vu},{Alishahiha:2008kc}}). So the central
charge and $\tilde{\Delta}$ of the CFT read
\be
c=12GQ^2l^4,\;\;\;\;\;\;\;\;{\tilde \Delta}=GQ^2l^4.
\ee
We would like
to identify this central charge as the central charge of a CFT whose global $SL(2,R)$ symmetry is
the isometry of the $AdS_2$ geometry. With the above particular value of
$k$, the CFT descriptions is consistent with gravity result. This 
is a strong indication in favor of the $AdS_2/CFT_1$ correspondence.

\subsection{Near horizon modes}

There is an alternative CFT living in the near horizon region of the black hole describing its entropy.
In this approach one may identify the entropy as the number of states at the horizon. To do this
we follow \cite{SOL} and make a change of variables in the action $S_1$,
\be 
\frac{e^{\phi}}{2G}=\Phi^2=q\Phi_0\varphi,\;\;\;\;\;\;\;g_{\mu\nu}\rightarrow e^{\frac{2}{q\Phi_0}\varphi} g_{\mu\nu},
\ee 
where $\Phi_0$ is the value of $\Phi$ at horizon, $\Phi_0^2=2GQ^2l^4$, and $q$ is a free
parameter and get 
\be
S_1=-\int d^2x\sqrt{g}\left(-\frac{1}{4}q\Phi_0\varphi R+\frac{1}{2}(\nabla \varphi)^2-\frac{2q\Phi_0}{l^2}\varphi e^{\frac{2}{q\Phi_0}\varphi}
+\frac{l^2}{32G}e^{-\frac{2}{q\Phi_0}\varphi} F^2\right).
\ee
Then integrating out the gauge field, using the Maxwell equations, leads to an effective potential 
for the scalar field $\varphi$ 
\be
S_1=-\int d^2x\sqrt{g}\left(-\frac{1}{4}q\Phi_0\varphi R+\frac{1}{2}(\nabla \varphi)^2+V(\varphi)\right).
\label{SOLAC}
\ee
This is the action considered in \cite{SOL} where it was shown that in the
near horizon limit ($r\rightarrow 0$) the energy momentum tensor of the theory becomes 
traceless leading to a CFT with a specific central charge. 

To be precise the author of \cite{SOL} starts from a $d$-dimensional spherically symmetric black hole and
decomposes the metric into two parts
\be
ds^2=-g(r)dt^2+\frac{dr^2}{g(r)}+h_{ij}(r)dx^idx^j.
\ee
Dimensionally reducing along the $x^i$'s he gets, after some manipulations, the above two 
dimensional action. 

Although the considerations of \cite{SOL} are for a non-extremal black hole where the leading 
behavior of $g$ in the near horizon limit is $g(r)\sim (r-r_h)$ with $r_h$ being the radius of the horizon,
the procedure works for the extremal case as well where $g(r)=v r^2$. The only difference is that in the
non-extremal case the trace of energy momentum tensor vanishes exponentially when 
the horizon is approached while in our case the approach is power law.

Using the change of variable $z=\frac{1}{r}$ (note that in this case horizon is at
$z\rightarrow \infty$), the components of the energy momentum tensor of the action (\ref{SOLAC}) read 
\bea
{T}_{00}&=&\frac{1}{4}((\partial_t\varphi)^2+(\partial_z\varphi)^2)-\frac{q\Phi_0}{4}(\partial_z^2\varphi-
\frac{v}{2z}\partial_z\phi)+\frac{v}{z^2}V(\varphi),\cr
{T}_{zz}&=&\frac{1}{4}((\partial_t\varphi)^2+(\partial_z\varphi)^2)+\frac{q\Phi_0}{4}(-\partial_t^2\varphi+
\frac{v}{2z}\partial_z\phi)-\frac{v}{z^2}V(\varphi),\cr
{T}_{0z}&=&\frac{1}{2}\partial_t\partial_z\phi-\frac{q\Phi_0}{4}(\partial_z\partial_t\varphi-\frac{v}{2z}
\partial_t\varphi).
\eea
Using the light-cone coordinates $z_{\pm}=t\pm z$ the non-zero components of the energy momentum
tensor are given by
\be
{T}_{\pm\pm}=(\partial_\pm\varphi)^2\mp\frac{1}{2} q\Phi_0\partial_\pm^2\varphi,
\ee
which has the same structure as (\ref{ee}). Essentially it has the form of a scalar field in the 
presence of a non-zero background charge.
If we define the Virasoro algebra as the Fourier coefficients in the expansion of the above 
energy momentum tensor, 
\be
L_n=\frac{L}{2\pi}\int_{-L/2}^{L/2}dz\;e^{2\pi i nz/L}T_{++},
\label{Vi}
\ee
then it is easy to see that the shifted Virasoro generators 
\be
\tilde{L}_n=L_n+\frac{c}{24}\delta_{n,0},
\ee
satisfy a Virasoro algebra with central charge $c=3\pi q^2\Phi_0^2$\cite{Cvitan:2002cs}. 
Here we compactified $z$ coordinate on a circle of circumference $L$. Finally we would like to send
$L$ to infinity. 

The aim is to evaluate $L_0$ for our background. It is, however, tricky as the scalar field 
is constant and naively leads to zero $L_0$. Nevertheless note that in the vicinity of the horizon 
the equations of motion allow a
more general zero mode configuration $\varphi=2\Phi_0/qL\;z $; in our notation the horizon is
at $z=L/2$. Plugging this expression in the definition of the energy momentum tensor and using
the definition of $L_n$, (\ref{Vi}), one arrives at
\be
c=6\pi q^2 G Q^2l^4,\;\;\;\;\;\;\;\;\tilde{\Delta}-\frac{c}{24}=\frac{GQ^2l^4}{\pi q^2}.
\ee
For $q^2=\frac{2}{\pi}$ these are exactly the same results we obtained using the
asymptotic symmetry of the model. 

Observe that the effect of the $U(1)$ gauge transformation at asymptotic boundary condition reflects, 
upon integrating out the gauge field, a non-zero background charge in the near horizon limit description of
the theory. Therefore in this case twisting at infinity has the same effect as having a non-zero background charge
at the horizon. 
We conclude that there may be a one to one correspondence between these two CFT's, one at infinity and 
the other at the horizon (for discussions on these two CFT's see for example \cite{Carlip:1999cy}).

\section{Type II 2D gravity}

In this section we consider 2D Maxwell-dilaton gravity based on the following action 
\be
S_2=\frac{1}{8G}\int d^2x\sqrt{-g}\; e^{\phi}\left(R+\frac{2}{l^2}-\frac{l^2}{4}e^{2\phi}F^2\right),
\label{S2}
\ee
which can actually be obtained from the 3D pure gravity with cosmological constant by 
reducing to two dimensions along an $S^1$. This action has been used to study
entropy of extremal black hole in the presence of higher order corrections
(see for example \cite{sen2}).

We will redo our previous section's computations for the action (\ref{S2}).
To start with note that this action also has an $AdS_2$ solution. 
Following our discussion in the previous section
we start with the following ansatz
\be
ds^2=v\left(-r^2 dt^2+\frac{dr^2}{r^2}\right),\;\;\;\;\;e^{\phi}=\eta,\;\;\;\;\;
F_{rt}=\frac{e}{l^2},
\ee
where $v,\eta$ and $e$ are constants to be determined by the equations of motion and use the
entropy function formalism to find
\be
ds^2=\frac{l^2}{4}(-r^2dt^2+\frac{dr^2}{r^2}),\;\;\;\;\;\;e^{\phi}=\sqrt{4GQl^2},\;\;\;\;\;\;
F_{tr}=\sqrt{\frac{1}{16GQl^2}}
\ee
with the entropy, 
\be
S_{BH}=2\pi\sqrt{\frac{Ql^2}{4G}}.
\label{ccc}
\ee 
Following our discussions in the previous section the goal is to reproduce this entropy using
the number of states of a CFT which can be defined by asymptotic symmetry of the $AdS_2$ solution.

Note that the above solution, unlike the solution in the previous section, has the
underlying  symmetry of $AdS_3$. 
In other words one may lift the two dimensional 
solution to three dimensions as follows
\be
ds_{(3)}^2=ds_{(2)}^2+l^2e^{2\phi}(dz+A_\mu dx^\mu)^2,
\ee   
which for our solution, defining $y=\sqrt{16GQl^2}z$, becomes
\be 
ds_{(3)}^2=\frac{l^2}{4}\left(dy^2+2r^2dy dt+\frac{dr^2}{r^2}\right),
\ee
clearly the $AdS_3$ written in the $S^1$ fibered over $AdS_2$ coordinates. So
we get the $AdS_3$ with an identification and the solutions is lifted
to the extremal BTZ black hole.

Taking into account the isometry of $AdS_3$ which is $SL(2,R)_R\times SL(2,R)_L$,  our 
two dimensional solution can be  thought of as an $SL(2,R)_L$ quotient of $AdS_3$ \cite{LS}. 
Therefore we are left with just the $SL(2,R)_R$ symmetry of $AdS_3$ which under the reduction  
reduces to the $SL(2,R)$ isometry of $AdS_2$
plus a gauge transformation \cite{Strominger:1998yg}. 
Following the arguments of \cite{HS} one can see that the conformal diffeomorphism in two dimensions
will not respect the boundary conditions of the gauge field, necessitating an extra 
$U(1)$ gauge transformation leading to a twisted energy momentum tensor \cite{Strominger:1998yg}
\be
\tilde{T}_{\pm\pm}=T_{\pm\pm}\pm\partial_\pm j_\pm,
\ee 
where $j_\pm$ is the appropriately normalized
 current associated with the Kaluza-Klein 
$U(1)$ gauge symmetry, and $T$ is the $AdS_2$ energy momentum tensor with 
zero central charge. 

The central charge of the twisted quantum theory crucially depends on the 
correct normalization of the $U(1)$ current. In our case there is a short 
cut in calculating the central charge. 
The Virasoro generators $\tilde{L}_n$ corresponding to the twisted energy momentum tensor $\tilde{T}$,
are equal to the right handed modes $L^{(3)}_n$ of the energy momentum tensor of $AdS_3$  corresponding to its 
$SL(2,R)_R$ isometry, $L_n^{(3)}=\tilde{L}_n$\cite{Strominger:1998yg}. As
\be
[L^{(3)}_n,L^{(3)}_m]=(n-m)L^{(3)}_{n+m}+\frac{c^{(3)}}{12} (n^3-n)\delta_{n+m,0},
\ee
with $c^{(3)}=3l/2G_3$, we get
\be
[\tilde{L}_n,\tilde{L}_m]=(n-m)\tilde{L}_{n+m}+\frac{1}{12}(\frac{3}{2G}) (n^3-n)\delta_{n+m,0},
\ee
where $G_3=lG$. This gives the central charge of the twisted energy momentum tensor, 
\be
c=\frac{3}{2G}.
\ee
The entropy can now be computed from the Cardy formula 
following the procedure of the section two and turns out to be consistent with the gravity computations
for $k=\frac{1}{4G}$ where $\tilde{\Delta}=Ql^2$. This again is a direct 
indication in favor of $AdS_2/CFT_1$ correspondence.

\section{Conclusions and discussions}

In this paper we have studied 2D Maxwell-dilaton gravity on $AdS_2$ geometry. We have seen that 
there are two distinctive models of 2D Maxwell-dilaton gravity defined by actions $S_1$ and 
$S_2$ in (\ref{actions}).

Both actions admit an $AdS_2$ vacuum solution. In both cases to maintain the
consistent boundary conditions of the fields in the theory the conformal diffeomorphism in the boundary at 
infinity has to be accompanied by a special $U(1)$ gauge transformation. 
The theories are well-described by the twisted energy momentum
tensor defined by
\be
\tilde{T}_{\pm\pm}=T_{\pm\pm}\pm A \partial_{\pm}j_{\pm},
\ee 
where $A$ is a constant depending on the normalization of the 
$U(1)$ current $j_{\pm}$. Although the Virasoro algebra of $T$ has zero
central charge, as expected from 2D quantum gravity, the twisted energy momentum 
tensor leads, after fixing the constant $A$ correctly, to the non-zero central charge given by
\be
c=12GQ^2l^4,\;\;\;\;\;\;\;c=\frac{3}{2G}
\ee
for the models based on $S_1$ and $S_2$, respectively; which can then be used in the Cardy formula to compute the
number of states of the CFT. It was then observed that the resultant entropy
is equal to the entropy of the $AdS_2$ solution obtained from gravity
calculations. This precise agreement can be thought of as a strong indication of
$AdS_2/CFT_1$ correspondence. In particular it might be a sign that the holographic
dual of gravity on $AdS_2$ is a chiral half 2D CFT. In fact one may go further and claim that
any consistent 2D quantum gravity on an $AdS_2$ is dual to a chiral half 2D CFT generalizing
the situation of $AdS_3$.

Although these models show certain similarities, they have significant differences. 
For example the central charge of the first
model depends on the detail of the solution considered, while in the second case
it only depends on the two dimensional Newton's constant.
In the first model even though we have a gauge field, the solution can not be lifted 
to an $AdS_3$ geometry. In fact there are  indications that the model may be
obtained from a higher dimensional extremal black hole. 
Actually following \cite{SOL} we started from a four dimensional extremal black hole 
with near horizon geometry $AdS_2\times M_2$ and reducing along 
the $M_2$ manifold we found an effective Maxwell-dilaton gravity on $AdS_2$. In this 
approach the entropy is associated with the near horizon modes where a new
CFT emerges. We have seen that the emergent CFT has the same central charge as that 
obtained using asymptotic symmetry. Moreover we have observed that the energy momentum tensor
has the form of a twisted energy momentum tenor where the extra twist term can be
interpreted as a non-zero background field (or Liouville theory). Indeed as far as the energy momentum
tensor, central charge and entropy are concerned there is a 
correspondence between the two CFT's; one at infinity and the other at the 
horizon. This observation requires further study.

On the other hand the $AdS_2$ solution of the second case was lifted to an $AdS_3$ solution, exhibiting
a direct connection between the two and three dimensional theories. In particular
we were able to relate the central charge of the twisted energy momentum tensor to
the Brown and Henneaux-like central charge.  

It is tempting to ask whether this connection can lead to an understanding of certain aspects of one theory
form the other. In particular it has been recently observed that by adding a Chern-Simons
term to the three dimensional gravity, a chiral gravity is obtained\cite{LSS}. It is then interesting 
to study the resultant chiral gravity in the 2D theory. Actually 
a two dimensional Chern-Simons term can be added to our theory. In our notation the Chern-Simons 
action is given by\cite{{Guralnik:2003we},{Grumiller:2003ad}}
\be
S_{cs}=\frac{1}{32G\mu}\int d^2x\;e^{2\phi}\left(lR\epsilon^{\mu\nu}F_{\mu\nu}+l^3e^{2\phi}
\epsilon^{\mu\nu}F_{\mu\rho}F^{\rho\delta} F_{\delta\nu}\right).
\ee
It is easy to show that a model based on $S_2+S_{cs}$ still has an $AdS_2$ solution. The
corresponding central charge can be computed and is seen to get
corrected. The correction of the central charge depends on the sign of the electric 
field $e$. Using entropy function formalism, we get
\bea
c&=&\frac{3}{2G}(1-\frac{1}{l\mu}),\;\;\;\;\;\;\;\;{\rm for}\;\;\;\;e>0\cr
c&=&\frac{3}{2G}(1+\frac{1}{l\mu}),\;\;\;\;\;\;\;\;{\rm for}\;\;\;\;e<0.
\eea
A chiral gravity in two dimensions, may be also defined in the same way as in 
three dimensions. We note, however, that a priori there is no reason from the two dimensional
point of view to force the relation $\mu l=1$. Nevertheless since the theory is related to the
three dimensional theory, one would expect to find some tachyonic modes leading  to a 
particular value for $\mu$. Work on this question  is in progress.

\vspace*{1cm}

{\bf Acknowledgments}

 This work is supported in part by 
Iranian TWAS chapter at ISMO. We would like to thank referee for his/her comments.

\end{document}